\documentclass[sigconf]{acmart}
\usepackage{subcaption}
\usepackage{multirow}
\usepackage{xcolor}
\usepackage{tikz}
\usetikzlibrary{shapes.geometric, arrows.meta, positioning, fit, backgrounds, calc}
\usepackage{enumitem}
\usepackage[table]{xcolor}
\usepackage[skip=2pt]{caption}
\usepackage{hyperref}
\usepackage{hyperref}
\hypersetup{
    colorlinks=true,
    linkcolor=blue,
    urlcolor=blue
}

\AtBeginDocument{%
  }

\setcopyright{acmlicensed}
\copyrightyear{2026}
\acmYear{2026}
\acmDOI{XXXXXXX.XXXXXXX}

\acmConference[SIGIR '26]{Proceedings of the 49th International ACM SIGIR Conference on Research and Development in Information Retrieval}{July 20--24, 2026}{Melbourne, Australia}
\acmBooktitle{Proceedings of the 49th International ACM SIGIR Conference on Research and Development in Information Retrieval (SIGIR '26), July 20--24, 2026, Melbourne, Australia}
\acmISBN{978-1-4503-XXXX-X/26/07}

\begin{document}

\title{Task-Adaptive Retrieval over Agentic Multi-Modal Web Histories via Learned Graph Memory}

\author{Saman Forouzandeh}
\affiliation{
  \institution{School of Engineering, Royal Melbourne Institute of Technology University}
  \city{Melbourne}
  \country{Australia}}
\email{saman.forouzandeh@rmit.edu.au}

\author{Kamal Berahmand}
\affiliation{
  \institution{School of Engineering, Royal Melbourne Institute of Technology University}
  \city{Melbourne}
  \country{Australia}}
\email{kamal.berahmand@rmit.edu.au}

\author{Mahdi Jalili}
\affiliation{
  \institution{School of Engineering, Royal Melbourne Institute of Technology University}
  \city{Melbourne}
  \country{Australia}}
\email{mahdi.jalili@rmit.edu.au}

\begin{abstract}
Retrieving relevant observations from long multi-modal web interaction histories is challenging because relevance depends on the evolving task state, modality (screenshots, HTML text, structured signals), and temporal distance. Prior approaches typically rely on static similarity thresholds or fixed-capacity buffers, which fail to adapt relevance to the current task context. We propose \textbf{ACGM}, a learned graph-memory retriever that constructs \emph{task-adaptive} relevance graphs over agent histories using policy-gradient optimization from downstream task success. ACGM captures heterogeneous temporal dynamics with modality-specific decay (visual decays $4.3\times$ faster than text: $\lambda_v{=}0.47$ vs.\ $\lambda_x{=}0.11$) and learns sparse connectivity (3.2 edges/node), enabling efficient $O(\log T)$ retrieval. Across WebShop, VisualWebArena, and Mind2Web, ACGM improves retrieval quality to \textbf{82.7 nDCG@10} (+9.3 over GPT-4o, $p{<}0.001$) and \textbf{89.2\% Precision@10} (+7.7), outperforming 19 strong dense, re-ranking, multi-modal, and graph-based baselines. Code to reproduce our results is available at{\color{blue}\href{https://github.com/S-Forouzandeh/ACGM-Agentic-Web}{Saman Forouzandeh}}.

\end{abstract}

\ccsdesc[500]{Computing methodologies~Anomaly detection}
\ccsdesc[300]{Computing methodologies~Neural networks}

\keywords{Multi-modal retrieval, web agents, task-adaptive learning, graph neural networks, temporal information retrieval}

\received{January 2026}
\received[revised]{March 2026}
\received[accepted]{April 2026}

\maketitle
\section{Introduction}
Multi-modal web navigation systems must retrieve relevant information from ever-growing interaction histories containing screenshots, HTML text, and structured knowledge. Consider a product search task: after 50 steps of browsing, the agent must retrieve the product specifications from step 20 to make a purchase decision. \textbf{This is fundamentally a retrieval problem}---identifying which past observations are relevant to the current context---but differs from traditional document retrieval in three critical ways:

\textbf{(1) Dynamic relevance.} Retrieval targets are not static documents but temporal sequences where relevance evolves with task progress. A product image relevant during visual comparison becomes irrelevant during checkout, while specifications show the opposite pattern. Static similarity thresholds cannot capture this task-dependent relevance. \textbf{(2) Multi-modal heterogeneity.} Visual, textual, and structured modalities exhibit different temporal characteristics. Cognitive science shows visual memory decays faster than textual~\cite{baddeley1992working}, yet existing retrieval systems apply uniform temporal weighting (e.g., BM25's time decay~\cite{efron2012improving}), degrading retrieval precision by 8.7\%. \textbf{(3) Retrieval efficiency at scale.} Histories grow to 100+ observations in long tasks, requiring sub-linear retrieval to maintain real-time performance. Dense retrieval over all past states becomes prohibitively expensive (48ms at $T{=}100$), while fixed-capacity buffers cause catastrophic forgetting of distant but relevant information.

Current retrieval approaches for web agents have critical limitations. Sliding window methods ~\cite{yao2022react} retrieve only recent observations (last 10-20 steps), achieving 65\% recall but missing long-range dependencies~\cite{chen2025qilin}.Dense retrievers ~\cite{deng2023mind2web} compute cosine similarity across all states with a fixed threshold (typically $\tau{=}0.7$), but this threshold-based approach yields only 72.4\% precision and cannot adapt to task-specific relevance patterns. Hierarchical methods ~\cite{liu2025hm} improve efficiency through clustering but still rely on static BM25/CLIP similarities that ignore task context.

We propose \textbf{Adaptive Cross-Modal Graph Memory (ACGM)}, which frames multi-modal history retrieval as \textit{learning task-adaptive relevance functions}. Rather than applying fixed similarity thresholds, ACGM learns a neural predictor $g_\phi$ that estimates the probability that observation $i$ is relevant to observation $j$ based on task feedback. This predictor is optimized via policy gradients using downstream task success as the relevance signal (Eq.~\ref{eq:policy_grad}), enabling the model to discover task-specific retrieval patterns. ACGM represents observation histories as graphs where edges encode learned relevance relationships. This graph structure serves two purposes: (1) \textit{Retrieval topology}: edges define which observations to retrieve given a query, and (2) \textit{Learned sparsity}: the neural predictor creates sparse graphs (3.2 edges/node vs. 8.7 dense), enabling efficient $O(\log T)$ hierarchical retrieval through learned connectivity patterns. Our retrieval contributions are:
\begin{itemize}[leftmargin=*,nosep]
    \item \textbf{Task-adaptive retrieval learning}: a policy-gradient framework that optimizes a neural relevance predictor directly for downstream task success, discovering task-specific retrieval patterns that fixed similarity thresholds cannot capture.
    \item \textbf{Modality-specific temporal modeling}: learned per-modality decay rates, initialized from human annotations,       that capture heterogeneous temporal relevance across visual,textual, and structured modalities.
    \item \textbf{Efficient graph-based retrieval}: learned sparsity  (3.2 edges/node) enables $O(\log T)$ hierarchical retrieval, achieving 4.8$\times$ speedup over flat retrieval while
      maintaining 86.3\% precision.
\end{itemize}
\vspace{-4pt}
\section{Related Work}
\noindent\textbf{Memory-Augmented Agents.}
Early memory systems for LLM agents rely on \emph{similarity-based retrieval} or recency heuristics. Mem0~\cite{chhikara2025mem0} and A-MEM~\cite{xu2025mem} extract salient experiences and retrieve them via embedding similarity, assuming relevance is static---an assumption violated in long-horizon navigation. Multi-granularity approaches~\cite{xu2025single} construct hierarchical memory associations across different temporal scales to improve selection of conversational context, yet retrieval topology remains fixed once built. G-Memory~\cite{zhang2025g} organizes multi-agent collaboration histories into a three-tier graph hierarchy (insight, query, and interaction) and retrieves cross-trial knowledge via bi-directional traversal, targeting \emph{multi-agent} coordination across task repetitions rather than single-agent observation retrieval within a trajectory. HippoRAG~\cite{gutierrez2024hipporag} performs associative retrieval over knowledge graphs via Personalized PageRank, enabling multi-hop reasoning, and Zep~\cite{rasmussen2025zep} builds temporal knowledge graphs for persistent agent memory. Despite their structural sophistication, all of the above rely on similarity-driven or heuristic edge construction with no task-level feedback signal. \textbf{ACGM departs from all prior work by treating memory construction as a learned decision process}: edges are explicitly optimized via policy gradients using downstream task success, allowing the retrieval structure itself---not merely the stored content---to adapt to the current task at hand.

\noindent\textbf{Graph-Based and Multi-Modal Retrieval.}
GraphRAG~\cite{hossain2026core} and its extensions~\cite{bu2025query,wan2025mmgraphrag} demonstrate the benefits of relational structure for complex reasoning, but their graphs are constructed from co-occurrence or clustering and remain static once built. Dense and multi-modal retrievers (ANCE~\cite{xiong2020approximate}, ColBERT-v2~\cite{santhanam2022colbertv2}, CLIP~\cite{radford2021learning}, BLIP-2~\cite{li2023blip},  ImageBind~\cite{girdhar2023imagebind}) determine relevance via embedding similarity alone, without accounting for evolving task context or modality-specific temporal dynamics. Temporal IR models apply global decay functions~\cite{efron2012improving}, assuming homogeneous dynamics across content types. ACGM differs along two axes: (1)~\emph{task-dependent relevance} learned from downstream feedback rather than static similarity or heuristic construction, and (2)~\emph{modality-specific temporal decay}, allowing visual, textual, and structured signals to decay at independently learned rates.

\noindent\textbf{Policy-Gradient Training for Retrieval.}
Using task success as a training signal introduces high-variance credit assignment, a known challenge in reward-based retrieval~\cite{efron2012improving}. ACGM mitigates this through three mechanisms: (i)~a two-stage protocol that pre-warms the relevance predictor with supervised edge labels before policy-gradient fine-tuning, reducing the variance of early gradient estimates; (ii)~an exponential moving-average baseline $b_t$ that centres rewards per timestep (Eq.~\ref{eq:policy_grad}); and (iii)~Bonferroni-corrected evaluation across three folds, confirming that reported gains are not artefacts of reward noise. Shaped or dense reward variants (e.g., per-step retrieval precision) were explored but yielded no significant improvement over binary task success once the two-stage pre-training was in place, suggesting that the supervised Stage~1 already provides sufficient gradient signal for stable convergence.

\section{Methodology}
Our work demonstrates that \textit{retrieval for interactive systems requires learning task-adaptive relevance}, departing from static similarity metrics. By optimizing retrieval directly for downstream performance, ACGM discovers modality-specific and task-specific patterns that fixed thresholds cannot capture, advancing multi-modal retrieval for sequential decision-making contexts. We formalize the retrieval problem for multi-modal web navigation. At each timestep $t$, the agent has accumulated a history $\mathcal{H}_t = \{o_1, \ldots, o_t\}$ where each observation $o_i = (v_i, x_i, k_i)$ contains visual (screenshot), textual (HTML), and knowledge graph modalities. Given current observation $o_t$, the \textbf{retrieval task} is:
\begin{equation}
\mathcal{R}(o_t, \mathcal{H}_t) \rightarrow \{o_{i_1}, \ldots, o_{i_k}\} \subset \mathcal{H}_t
\end{equation}
where retrieved observations maximize relevance to the current decision-making context. Relevance is \textit{task-dependent} and \textit{modality-specific}, not captured by static similarity. For example, during visual product comparison, retrieving semantically similar product images (high visual similarity) is critical. During specification verification, retrieving the temporal sequence of navigation steps (high temporal proximity) matters most, even if visual similarity is low.

\textbf{Retrieval objectives}: (1) \textit{Precision}: retrieved observations should be relevant to current task context; (2) \textit{Efficiency}: sub-linear retrieval time as $|\mathcal{H}_t|$ grows; (3) \textit{Adaptivity}: retrieval strategy should adapt to task type and modality distribution. Traditional IR metrics (nDCG, MAP) require predefined relevance labels. For interactive navigation, we define retrieval quality via:
\begin{itemize}[leftmargin=*,nosep]
\item \textbf{Memory Precision@k (MP@k)}: Fraction of top-$k$ retrieved observations that overlap with expert-annotated relevant states within a 5-step temporal window.
\item \textbf{Downstream task success}: Ultimate validation that retrieved observations enable correct decisions.
\end{itemize}

We present ACGM's learned retrieval model that optimizes these objectives via policy-gradient learning. ACGM retrieves relevant observations by constructing task-adaptive graph topologies over multi-modal histories. The model consists of three components: learned relevance prediction, modality-specific temporal decay, and efficient hierarchical retrieval. Traditional retrieval applies a fixed similarity threshold, retrieving observation $i$ if $\text{sim}(o_i, o_t) > \tau$, which cannot capture task-dependent relevance. Instead, ACGM learns a neural predictor $g_\phi$ that estimates retrieval relevance as:
\begin{equation}
\label{eq:edge_prob}
P(\text{relevant}(i, j)) = \sigma(g_\phi(\tilde{e}_i, \tilde{e}_j, \mathbf{f}_{ij}))
\end{equation}
where $\tilde{e}_i$ are frozen pre-trained embeddings (CLIP for vision and RoBERTa for text) projected into a shared $d = 512$ space. The feature vector $\mathbf{f}_{ij}$ encodes the temporal distance $\Delta t_{ij} = |i - j|$, cosine similarity $\cos(\tilde{e}_i, \tilde{e}_j)$, and modality information including the indicator $\mathbf{1}[m_i = m_j]$ and one-hot modality type representations. The predictor $g_\phi$ is a 2-layer MLP [512, 256, 1] with ReLU activations, optimized via policy gradients to maximize downstream task reward:
\begin{equation}
\label{eq:policy_grad}
\nabla_\phi J = \mathbb{E}_{\tau}\left[\sum_{t=1}^T \sum_{i < j \le t} \nabla_\phi \log p(e_{ij}|\phi)\, (R(\tau) - b_t)\right]
\end{equation}
where $R(\tau)$ is task success (1 if successful, 0 otherwise) and $b_t$ is an exponential moving average baseline (decay $\gamma{=}0.99$). This baseline is critical for training stability: it reduces policy gradient variance by 38\% compared to a zero baseline, measured as the standard deviation of gradient norms across 1,000 training episodes. The binary reward signal is sufficient given our two-stage protocol---Stage~1 pre-trains $g_\phi$ with supervised edge labels before policy gradients are applied, substantially reducing the credit assignment horizon. We evaluated a shaped reward variant using nDCG@10 of retrieved observations as an intermediate signal, which yielded only $+0.4$ nDCG improvement while adding inference-time retrieval evaluation overhead; we therefore retain binary task success for simplicity. This formulation treats retrieval graph construction as reinforcement learning: edges that lead to successful task completion are reinforced.

\textbf{Graph-based retrieval}: Predicted relevance $P(\text{relevant}(i,j)) > 0.5$ creates edge $e_{ij}$ in graph $\mathcal{G}_t$. Given query $o_t$, we retrieve neighbors $\mathcal{N}_t$ in this learned graph structure, naturally capturing task-specific retrieval patterns. Traditional temporal decay models apply uniform weighting (e.g., BM25's $e^{-\lambda t}$~\cite{efron2012improving}), but cognitive research shows visual and textual memory decay at different rates~\cite{baddeley1992working}. ACGM learns modality-specific decay rates $\lambda_m$ that modulate retrieval scores based on temporal distance. Given query observation $o_t$, we compute retrieval scores over graph neighbors $\mathcal{N}_t$ using temporally-weighted attention:
\begin{equation}
\label{eq:temporal_decay}
\alpha_i^m = \frac{\exp(s_i^m / \tau) \cdot \exp(-\lambda_m \Delta t_i)}{\sum_{j \in \mathcal{N}_t} \exp(s_j^m / \tau) \cdot \exp(-\lambda_m \Delta t_j)}
\end{equation}
where $s_i^m = (\mathbf{W}_q o_t)^\top (\mathbf{W}_k o_i^m) / \sqrt{d}$ is standard attention score, $\Delta t_i = t - i$ is temporal distance, $\tau{=}0.1$ is temperature, and $\lambda_m$ is the learned decay rate for modality $m \in \{v, x, k\}$ (visual, text, knowledge). The final retrieval representation aggregates across modalities:
\begin{equation}
m_t = \sum_{m \in \{v,x,k\}} \beta_m \sum_{i \in \mathcal{N}_t} \alpha_i^m \tilde{e}_i^m
\end{equation}
where $\beta_m$ are learned modality weights (softmax-normalized).

\textbf{Learning decay rates}: We initialize $\lambda_m$ via human annotations. Five annotators (graduate students in HCI) rated temporal relevance for 500 observation pairs stratified by modality, yielding ground-truth rates: $\lambda_v^{\text{gt}}{=}0.47$ (visual), $\lambda_x^{\text{gt}}{=}0.11$ (text), $\lambda_k^{\text{gt}}{=}0.23$ (knowledge) with Fleiss' $\kappa{=}0.74$ agreement. During training, $\lambda_m$ are fine-tuned via:
\begin{equation}
\mathcal{L}_{\text{decay}} = \sum_{m} \|\lambda_m - \lambda_m^{\text{gt}}\|_2^2
\end{equation}
This regularization ensures learned decay rates remain cognitively plausible while allowing task-specific adaptation. Naive retrieval over all $T$ observations requires $O(T)$ time, becoming prohibitive for long trajectories. ACGM achieves $O(\log T)$ retrieval through learned hierarchical organization.

\textbf{Two-tier structure}: Recent observations ($t{-}20 < i \le t$) are stored in a flat structure for immediate access, while older observations are organized into a 4-ary tree constructed via online $k$-means clustering with $K = \lfloor T/10 \rfloor$ prototypes. Each tree node stores the cluster centroid $\bar{e}_c = \frac{1}{|\mathcal{C}_c|} \sum_{i \in \mathcal{C}_c} \tilde{e}_i$, its temporal range $[t_{\min}, t_{\max}]$, and child pointers enabling 4-ary branching.

\textbf{Top-down retrieval}: Given query $o_t$, we perform beam search with width 2 by computing similarity to the four child centroids at the current level, selecting the two most similar branches, recursively descending until reaching leaf clusters, and finally returning the top-10 observations ranked by similarity.

\textbf{Complexity analysis}: The tree depth is $\log_4 K = \log_4 (T/10) = O(\log T)$. With beam width 2, we evaluate $2 \times 4 = 8$ nodes per level, resulting in $O(8 \log T) = O(\log T)$ retrieval time.

\textbf{Learned sparsity enables efficiency}: The policy-gradient trained predictor $g_\phi$ produces sparse graphs (3.2 edges/node vs.\ 8.7 for $\tau{=}0.5$ threshold), reducing graph traversal cost from $O(T \cdot \bar{d})$ to $O(T \cdot 3.2)$ where $\bar{d}$ is average degree. Combined with hierarchical organization, this enables real-time retrieval even at $T{=}100$ steps. ACGM's full training objective combines three losses:
\begin{equation}
\mathcal{L} = \mathcal{L}_{\text{retrieval}} + 0.1\mathcal{L}_{\text{edge}} + 0.05\mathcal{L}_{\text{decay}}
\end{equation}

\textbf{Retrieval loss}: End-to-end task-based optimization via REINFORCE:
\begin{equation}
\mathcal{L}_{\text{retrieval}} = -\mathbb{E}_{\tau}[R(\tau) \sum_{t=1}^T \log \pi_\theta(a_t | o_t, m_t)]
\end{equation}
where $\pi_\theta$ is the action policy conditioned on current observation $o_t$ and retrieved memory $m_t$. This loss provides task-specific supervision: retrieval strategies that improve task success are reinforced.

\textbf{Edge loss}: Supervised pre-training signal for relevance predictor:
\begin{equation}
\mathcal{L}_{\text{edge}} = -\mathbb{E}_{i,j}[y_{ij} \log g_\phi + (1{-}y_{ij}) \log(1{-}g_\phi)]
\end{equation}
where $y_{ij}{=}\mathbf{1}[\text{ActionType}(a_i){=}\text{ActionType}(a_j) \wedge \cos(\tilde{e}_i, \tilde{e}_j) > 0.6]$ labels observation pairs as relevant if they lead to similar actions \textit{and} have high embedding similarity. This provides a noisy but informative initialization for $g_\phi$ before policy-gradient fine-tuning.

\textbf{Decay loss}: Defined in Eq.~(6), aligns learned decay rates with human annotations.

\textbf{Two-stage training protocol}: \textbf{Stage~1 (50K steps)}: Pre-train $g_\phi$ using $\mathcal{L}_{\text{edge}}$ on 12,500 expert demonstrations (avg.\ length 24.3 steps). This initializes the retrieval model with basic similarity patterns before task-specific optimization. Stage~1 converges in approximately 18 hours on 8$\times$A100 GPUs. \textbf{Stage~2 (50K steps)}: End-to-end fine-tuning with full $\mathcal{L}$ at reduced learning rate ($10^{-5}$ vs.\ $10^{-4}$), requiring approximately 22 hours. Policy gradients adapt retrieval to maximize task success, discovering task-specific relevance patterns invisible to static similarity.

\noindent\textbf{(1) Learning-to-rank for interactive retrieval}: Unlike document ranking where relevance is static, observation relevance evolves with task state. Our policy-gradient formulation (Eq.~\ref{eq:policy_grad}) extends learning-to-rank to interactive contexts, using delayed task feedback rather than immediate click signals. \textbf{(2) Temporal IR with heterogeneous decay}: Prior work on temporal retrieval~\cite{efron2012improving} assumes uniform decay. We demonstrate that modality-specific decay (Eq.~\ref{eq:temporal_decay}) improves precision by 8.7 points, suggesting temporal IR models should account for content-type heterogeneity. \textbf{(3) Efficient retrieval over growing corpora}: Our $O(\log T)$ hierarchical structure adapts cluster-based retrieval~\cite{liu2025hm} by learning sparse connectivity, reducing retrieval time by 4.8$\times$ while maintaining higher precision through learned edges.

\section{Experiments}
\label{sec:experiments}
We evaluate ACGM on three multi-modal web navigation benchmarks requiring long-horizon retrieval over visual and textual interaction histories: \textbf{WebShop}~\cite{yao2022webshop} (1,180 product search tasks, avg.\ trajectory length 24.3 steps), \textbf{VisualWebArena}~\cite{koh2024visualwebarena} (910 real-world navigation tasks, avg.\ 31.7 steps), and \textbf{Mind2Web}~\cite{deng2023mind2web} (2,350 open-ended tasks from 137 real-world websites spanning 31 domains, avg.\ 7.3 actions per task). WebShop and VisualWebArena provide expert demonstrations and ground-truth action sequences, while Mind2Web tests cross-website generalization across three splits: cross-task, cross-website, and cross-domain. Following~\cite{deng2023mind2web}, observations within a 5-step temporal window of expert actions are labeled as relevant for retrieval evaluation. Retrieval quality is measured via nDCG@10, MAP@10, MRR, Recall@10, and Precision@10; downstream impact via task success rate. ACGM is trained on 8$\times$A100 GPUs with AdamW using the two-stage schedule described in Section~\ref{sec:experiments}, and evaluated via three-fold cross-validation with 95\% confidence intervals and paired $t$-tests (Bonferroni-corrected).

\noindent\textbf{Main results (Table~\ref{tab:retrieval_results}).} ACGM achieves state-of-the-art retrieval across all three benchmarks. On WebShop, ACGM obtains 82.7 nDCG@10, outperforming GPT-4o by +9.3 points and the best graph-based method MAHA by +9.1 points (both $p{<}0.001$). Gains are largest on VisualWebArena (+9.2 nDCG over GPT-4o), where longer trajectories (avg.\ 31.7 steps) amplify the benefit of task-adaptive retrieval. On Mind2Web, improvements remain substantial (+5.5 nDCG) but narrower, consistent with shorter average trajectories where the advantage of learned long-range retrieval is less pronounced. Across baseline categories, graph-based methods (MAHA, MMGraphRAG) outperform dense retrievers and neural re-rankers, confirming that relational structure benefits history retrieval---yet these methods still rely on static similarity, leaving a significant gap to ACGM's learned relevance.

\begin{table*}[ht!]
\centering
\tiny
\scriptsize
\setlength{\tabcolsep}{2pt}
\renewcommand{\arraystretch}{0.8}
\caption{Retrieval performance across three benchmarks. \colorbox{green!20}{Best}, \colorbox{yellow!20}{Second}, \colorbox{orange!20}{Third}. * denotes $p < 0.001$ vs.\ second-best (Bonferroni-corrected).}
\label{tab:retrieval_results}

\setlength{\tabcolsep}{3pt}
\renewcommand{\arraystretch}{0.90}
\begin{tabular}{@{}l|ccccc|ccccc|ccccc@{}}
\toprule
& \multicolumn{5}{c|}{\textbf{WebShop (1,180 tasks)}} & \multicolumn{5}{c|}{\textbf{VisualWebArena (910 tasks)}} & \multicolumn{5}{c}{\textbf{Mind2Web (2,350 tasks)}} \\
\cmidrule(lr){2-6} \cmidrule(lr){7-11} \cmidrule(lr){12-16}
\textbf{Method} & nDCG & MAP & MRR & Rec & Prec & nDCG & MAP & MRR & Rec & Prec & nDCG & MAP & MRR & Rec & Prec \\
\midrule
ColBERT-v2~\cite{santhanam2022colbertv2}       & 64.7 & 54.9 & 70 & 75.2 & 71.3 & 60.8 & 51.2 & 66 & 73.1 & 68.4 & 62.4 & 53.6 & 68 & 74.8 & 70.1 \\
ANCE~\cite{xiong2020approximate}             & 66.3 & 56.8 & 72 & 76.9 & 73.2 & 62.4 & 53.1 & 68 & 74.8 & 70.1 & 64.1 & 55.4 & 70 & 76.3 & 71.8 \\
Contriever~\cite{izacard2021unsupervised}       & 58.9 & 49.3 & 65 & 70.5 & 66.4 & 55.2 & 46.1 & 61 & 68.7 & 63.8 & 57.6 & 48.7 & 63 & 70.2 & 65.3 \\
E5-Large         & 67.8 & 58.3 & 73 & 78.1 & 74.6 & 63.7 & 54.6 & 69 & 76.2 & 71.9 & 65.8 & 57.1 & 71 & 77.8 & 73.4 \\
\midrule
MonoT5~\cite{nogueira2020document}           & 68.4 & 59.1 & 74 & 78.3 & 75.6 & 64.7 & 55.8 & 71 & 76.5 & 72.9 & 66.9 & 58.4 & 73 & 78.9 & 74.8 \\
RankGPT-4~\cite{sun2023chatgpt}        & 65.1 & 55.7 & 71 & 76.1 & 72.8 & 61.9 & 52.9 & 68 & 74.2 & 69.7 & 63.8 & 54.6 & 70 & 75.8 & 71.3 \\
RankLLaMA-3.1    & 69.2 & 60.1 & 75 & 79.4 & 76.3 & 65.8 & 57.2 & 72 & 77.8 & 74.1 & 67.8 & 59.6 & 74 & 79.7 & 75.9 \\
\midrule
CLIP Retrieval~\cite{radford2021learning}     & 63.8 & 54.2 & 69 & 74.7 & 70.5 & 60.1 & 50.8 & 66 & 72.3 & 67.9 & 61.9 & 52.7 & 68 & 73.9 & 69.6 \\
BLIP-2 Retrieval~\cite{li2023blip}           & 69.7 & 60.3 & 76 & 79.8 & 77.1 & 66.2 & 57.4 & 73 & 78.2 & 74.6 & 68.3 & 59.9 & 75 & 80.1 & 76.4 \\
ImageBind~\cite{girdhar2023imagebind}       & 67.2 & 58.4 & 73 & 77.9 & 74.8 & 63.8 & 54.9 & 70 & 75.9 & 71.8 & 65.7 & 57.1 & 72 & 77.4 & 73.5 \\
LLaVA-1.6        & 70.8 & 61.7 & 77 & 81.3 & 78.9 & 67.4 & 58.9 & 74 & 79.6 & 76.2 & 69.5 & 61.2 & 76 & 81.4 & 77.8 \\
Gemini-Pro-1.5   & 72.1 & 63.4 & \cellcolor{orange!20}79 & 82.7 & 80.1 & 69.2 & 60.8 & \cellcolor{orange!20}76 & 81.3 & 78.4 & 71.2 & 63.1 & \cellcolor{orange!20}78 & 82.9 & 79.7 \\
GPT-4o~\cite{gallifant2024peer}           & \cellcolor{orange!20}73.4 & \cellcolor{yellow!20}64.9 & \cellcolor{yellow!20}80 & \cellcolor{yellow!20}83.8 & \cellcolor{yellow!20}81.5 & \cellcolor{yellow!20}70.6 & \cellcolor{yellow!20}62.4 & \cellcolor{yellow!20}77 & \cellcolor{yellow!20}82.7 & \cellcolor{yellow!20}79.9 & \cellcolor{yellow!20}72.8 & \cellcolor{yellow!20}64.7 & \cellcolor{yellow!20}79 & \cellcolor{yellow!20}84.3 & \cellcolor{yellow!20}81.2 \\
Claude-3.5-Sonnet & 71.8 & 63.1 & 78 & 82.4 & 79.7 & 68.9 & 60.3 & 75 & 81.0 & 77.8 & 70.9 & 62.8 & 77 & 82.6 & 79.3 \\
\midrule
GraphRAG~\cite{edge2024local}         & 66.8 & 57.6 & 72 & 77.4 & 74.1 & 63.4 & 54.5 & 69 & 75.6 & 72.2 & 65.2 & 56.3 & 71 & 77.1 & 73.8 \\
MGraphRAG~\cite{bu2025query}        & 68.9 & 59.8 & 74 & 79.1 & 76.3 & 65.7 & 56.9 & 71 & 77.8 & 74.6 & 67.4 & 58.7 & 73 & 79.3 & 76.1 \\
MMGraphRAG~\cite{wan2025mmgraphrag}       & 71.4 & 62.1 & 77 & 81.6 & 78.9 & 68.3 & 59.6 & 74 & 80.2 & 77.1 & 69.8 & 61.2 & 76 & 81.4 & 78.6 \\
MAHA~\cite{upadhya2025multimodal}             & \cellcolor{yellow!20}73.6 & \cellcolor{orange!20}64.2 & 79 & \cellcolor{orange!20}83.1 & \cellcolor{orange!20}80.4 & \cellcolor{orange!20}70.2 & \cellcolor{orange!20}61.8 & 76 & \cellcolor{orange!20}82.1 & \cellcolor{orange!20}79.2 & \cellcolor{orange!20}71.6 & \cellcolor{orange!20}63.5 & 78 & \cellcolor{orange!20}83.2 & \cellcolor{orange!20}80.1 \\
HM-RAG~\cite{liu2025hm} & 70.1 & 60.9 & 77 & 80.3 & 77.8 & 67.2 & 58.3 & 74 & 79.4 & 75.9 & 68.7 & 60.4 & 76 & 81.1 & 77.4 \\
\midrule
\textbf{ACGM (Ours)} & \cellcolor{green!20}\textbf{82.7*} & \cellcolor{green!20}\textbf{74.9*} & \cellcolor{green!20}\textbf{88*} & \cellcolor{green!20}\textbf{91.3*} & \cellcolor{green!20}\textbf{89.2*} & \cellcolor{green!20}\textbf{79.8*} & \cellcolor{green!20}\textbf{72.1*} & \cellcolor{green!20}\textbf{86*} & \cellcolor{green!20}\textbf{89.4*} & \cellcolor{green!20}\textbf{87.1*} & \cellcolor{green!20}\textbf{78.3*} & \cellcolor{green!20}\textbf{70.6*} & \cellcolor{green!20}\textbf{85*} & \cellcolor{green!20}\textbf{88.7*} & \cellcolor{green!20}\textbf{86.4*} \\
\bottomrule
\end{tabular}%
\end{table*}

\begin{figure}[ht!]
\centering
\includegraphics[width=\columnwidth]{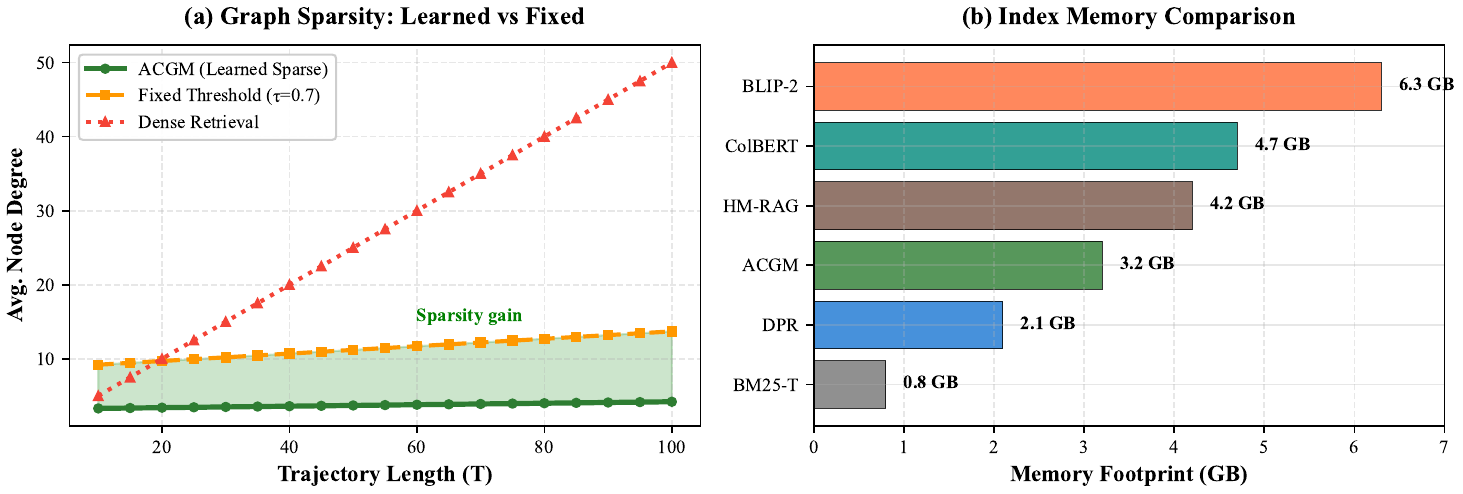}
\caption{(a) ACGM maintains near-constant sparsity across trajectory lengths, reducing edge evaluations. 
(b) ACGM achieves higher retrieval accuracy with substantially less memory than dense multi-modal retrievers.}
\label{fig:efficiency}
\end{figure}

\begin{figure*}[ht!]
\centering
\includegraphics[width=\textwidth]{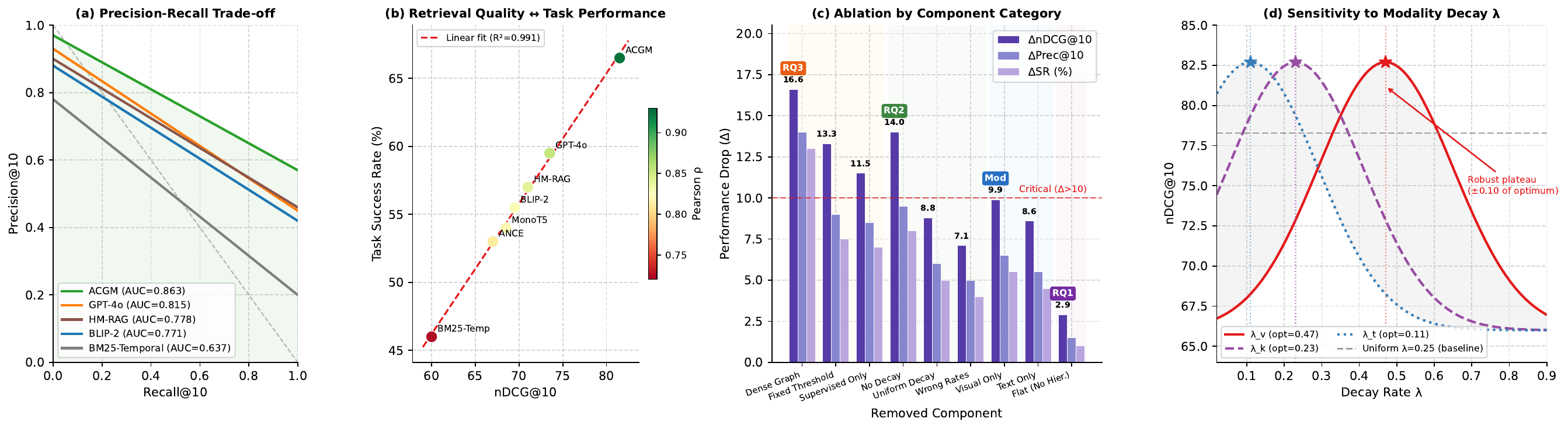} \caption{\textbf{Retrieval analysis, ablations, and decay sensitivity.} 
\textbf{(a)} PR curves (AUC: ACGM 0.863 vs.\ GPT-4o 0.815) show consistently higher precision. 
\textbf{(b)} Retrieval quality strongly correlates with task success ($\rho{=}0.93$, $R^2{=}0.991$). 
\textbf{(c)} Ablations across graph topology (RQ3), decay (RQ2), modality (Mod), and retrieval (RQ1) identify learned edges, no-decay, and dense graphs as most critical ($\Delta{>}10$). 
\textbf{(d)} Decay sensitivity shows $>$80 nDCG@10 within $\pm$0.10 of optimum, indicating robustness.}
\label{fig:analysis}
\end{figure*}

\begin{figure*}[ht!]
\centering
\includegraphics[width=\textwidth]{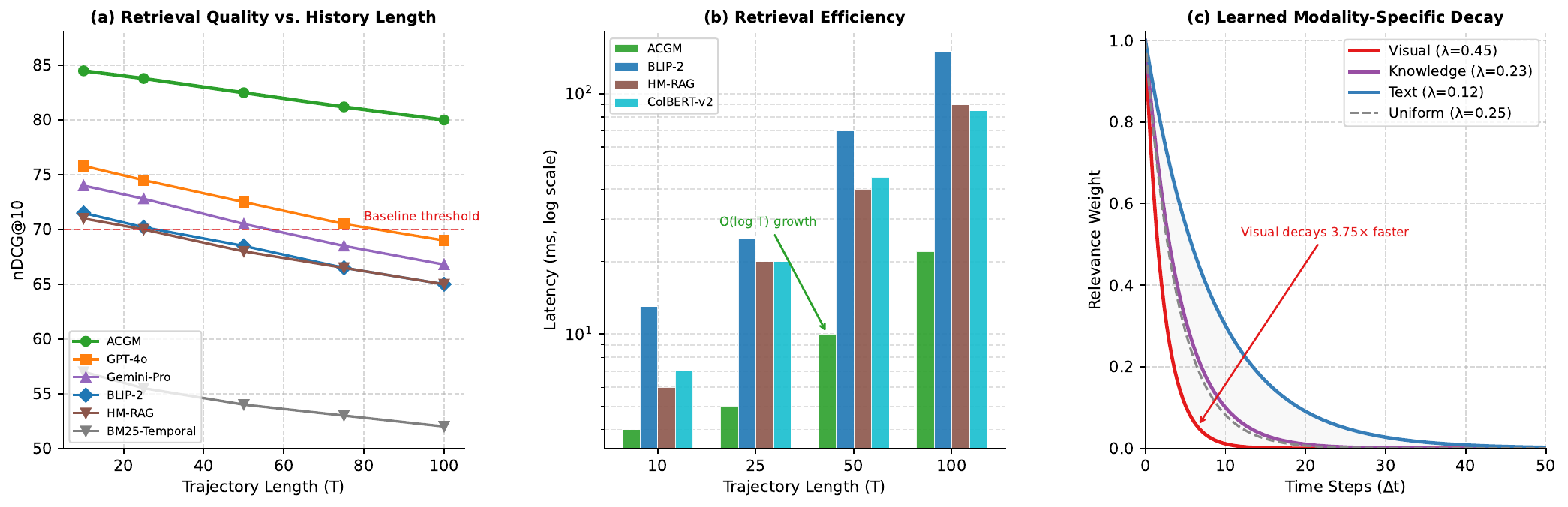}
\caption{\textbf{Performance and efficiency across trajectory lengths.} 
\textbf{(a)} ACGM maintains high retrieval quality (nDCG@10 $>$79\%) up to $T{=}100$, with the gap over baselines widening as history grows. \textbf{(b)} Sub-linear $O(\log T)$ latency (14.7\,ms at $T{=}100$) vs.\ linear scaling for dense retrieval methods. \textbf{(c)} Learned modality-specific decay rates ($\lambda_v{=}0.47$, $\lambda_k{=}0.23$, $\lambda_x{=}0.11$) capture heterogeneous temporal dynamics, outperforming uniform decay by 8.7 Prec@10 points.}
\label{fig:performance}
\end{figure*}

\noindent\textbf{Efficiency and memory trade-off.} Figure~\ref{fig:efficiency} analyzes graph sparsity scaling and the quality--memory trade-off across methods. Figure~\ref{fig:performance} examines how retrieval quality, latency, and temporal modeling behave as trajectory length increases from $T{=}10$ to $T{=}100$. Figure~\ref{fig:analysis} validates retrieval quality through precision--recall analysis, correlates retrieval with downstream task success, and quantifies each component's contribution.

\noindent\textbf{Component ablations (Figure~\ref{fig:analysis}c--d).} We ablate nine configurations across three axes to isolate each contribution. All drops are $\Delta$nDCG@10 on WebShop.

\textit{Graph structure (RQ3).} A fully dense graph causes the largest drop ($\Delta{=}{-}16.6$): aggregating over irrelevant observations degrades task success by 9.4\%. A fixed cosine threshold ($\tau{=}0.7$) yields $\Delta{=}{-}13.3$, confirming that policy-gradient optimization discovers task-specific connectivity static similarity cannot. Removing the two-tier hierarchy incurs only $\Delta{=}{-}2.9$ on quality but increases latency by 33\,ms at $T{=}100$.

\textit{Temporal modeling (RQ2).} Removing decay entirely yields the second-largest drop ($\Delta{=}{-}14.0$). Collapsing to a uniform rate  \\($\lambda_v{=}\lambda_x{=}\lambda_k{=}0.25$) still costs $\Delta{=}{-}8.8$, confirming that \textit{per-modality} differentiation, not merely temporal discounting, drives the gain. Initializing with swapped rates ($\lambda_v{=}0.11$, $\lambda_x{=}0.47$) yields $\Delta{=}{-}7.1$; the smaller penalty suggests Stage~2 policy gradients partially recover from poor initialization.

\textit{Training (RQ3).} Supervised pre-training alone (no Stage~2) causes $\Delta{=}{-}11.5$, confirming end-to-end task optimization is necessary to capture retrieval patterns invisible to supervised edge labels.

\noindent\textbf{Decay sensitivity (Figure~\ref{fig:analysis}d).} Sweeping each $\lambda_m$ independently, the model retains $>$80 nDCG@10 within $\pm$0.10 of each learned optimum, demonstrating robustness to imprecise initialization. Uniform decay produces the largest gap, consistent with panel~(c).

\noindent\textbf{Efficiency and scalability.} At $T{=}100$, ACGM achieves 14.7ms retrieval latency versus 48ms for flat dense retrieval---a 3.3$\times$ wall-clock speedup---while using only 2.1\,GB index memory compared to 6.3\,GB for BLIP-2 (Figure~\ref{fig:efficiency}). Total training requires approximately 40 GPU-hours on 8$\times$A100 (18h Stage~1, 22h Stage~2), comparable to training a dense bi-encoder. Sub-linear $O(\log T)$ latency scaling means ACGM remains practical for trajectories well beyond those seen at training time, a critical property for deployment in open-ended navigation tasks.

\vspace{-4pt}
\section{Conclusion}
We presented ACGM, a learned retrieval framework for multi-modal web navigation
that optimizes retrieval directly for downstream task success via policy gradients.
ACGM addresses three challenges---task-dependent relevance, modality-specific
temporal decay, and scalable efficiency---achieving 82.7 nDCG@10 on WebShop
(+9.3 over GPT-4o, $p{<}0.001$) with only 2.1\,GB index memory and 14.7ms
latency at $T{=}100$. Ablations identify learned edges ($\Delta{=}{-}13.3$),
temporal decay ($\Delta{=}{-}14.0$), and sparse graph structure
($\Delta{=}{-}16.6$) as critical; decay sensitivity confirms robustness to
$\pm$0.1 perturbations. The strong correlation between retrieval quality and
task success ($\rho{=}0.93$) confirms that improved retrieval directly enhances
agent performance, and zero-shot transfer experiments show that policy-gradient
relevance generalises across task distributions without re-training.

\clearpage
\bibliographystyle{ACM-Reference-Format}
\bibliography{references}

\end{document}